# The history of the Russian study of Chinese astronomy

Galina I. Sinkevich

Saint Petersburg State University of Architecture and Civil Engineering

galina.sinkevich@gmail.com

*Abstract*. The first relations between Russia and China date to the 13th century. From the 16th c., Russia sent ambassadors to China, who made a description of the country. In the 17th century, a Russian Orthodox mission was founded, led by Father Maxim Tolstoukhov. In the 18th century, the scholars of St.-Petersburg Academy of Sciences took a great interest in the history of Chinese astronomy in letters sent by the Jesuit mission in China. In the 19th c., the Orthodox mission in China began to carry out many functions – trade, diplomacy, science. Petersburg academy sent students to study various aspects of life in China, laying the foundations of Russian sinology. In 1848, St-Petersburg academy founded in Beijing a magnetic and meteorological observatory headed by K. Skachkov. He lived in China for 25 years and made extensive studies of the history of Chinese astronomy. He not only mastered Chinese but also studied many old manuscripts on astronomy. He wrote the research "The fate of astronomy in China" (1874). After Skachkov, Chinese astronomy was studied by G.N. Popov (1920), A.V. Marakuev (1934), and E.I. Beryozkina, who translated "Mathematics in nine books" into Russian (1957) and published a monograph on the history of Chinese mathematics, 1980. In 1995-2003, Beryozkina's post-graduate student at Moscow University, Fang Yao (Beijing), made a partial translation of the "Treatise on the Gnomon" (Zhoubi Suanjing) into Russian. In 2009, the five-volume encyclopedia "The Spiritual Culture of China" was published, with detailed articles by V.Ye. Yeremeev on the history of science, including astronomy.

Russians settled in China in the late 17[th] century. The first Orthodox mission was opened in 1685, when after a military conflict in the Amur region, a garrison of the Russian fortress of Albazin was captured, and 45 Cossacks were taken to Beijing. The Kangxi emperor provided them with a Buddhist shrine, which they converted to a chapel in honor of St. Nicholas. Among them was the Orthodox priest Father Maxim Tolstoukhov (died ca. 1711). All the settlers were included in the hereditary military caste, who under the laws of China were in second place after civil servants. On an equal basis with other soldiers, they received wages, and they were also given wives from the criminal order – widows of executed criminals. From this time, extensive diplomatic and trade relations between the nations began in Beijing. Emperor Peter the Great and the Holy Synod highly valued and respected Father Maxim for his successful mission in Beijing. Peter wrote to the head of the Siberian department: "Your worship writes that in Beijing Christians have built a church of our faith, and that many Chinese have been christened: this is a wonderful achievement. But for God's sake, be cautious there, to avoid drawing the ire of the Chinese authorities, and of the Jesuits, who are long established there. Therefore, we need priests who are not so much learned as they are wise and humble, so that arrogance does not cause this holy work to fail, as was the case in Japan" [Ustryalov]. In 1713-1716 the Russian religious mission was founded in China, which besides providing spiritual guidance for Orthodox believers in Beijing, had the task of scientific activity and diplomatic functions.

In St. Petersburg[1], the Imperial Petersburg academy of sciences regularly published reports by the Beijing Jesuit mission, including 342 letters from Father Antoine Gaubil (1689–

---

[1] "Novi commentarii Academiae Petropolitanae", 1754

1759) on the advanced state of Chinese astronomy, and maps of China. These were read by Leonard Euler.

Members of the Russian religious mission studied Chinese and Manchurian, and the history, culture and religion of China. Among the secular members of the mission, young people were appointed from students of educational institutions and the religious academy. The Imperial St. Petersburg academy of science also made active use of the mission to carry out its own scientific programs. The Chinese authorities were tolerant of the academy. The students studied medicine, mathematics, literature and philosophy, the system of Confucius, the history, geography, statistics and jurisprudence of the Chinese state. Unfortunately, they did not take an interest in Chinese astronomy. The Beijing mission gave Russia its first prominent Sinologists: I.K. Rassokhin, A.L. Leontiev, I.Ya Bichurin[2], O.M. Kovalevsky, I.P. Voitsekhovsky, I.I. Zakharov, P.I. Kafarov, V.P. Vasiliev. Their scientific legacy formed the basis of Russian Sinology. In 1861 the Russian diplomatic mission was founded in Beijing [Feklova].

In 1819, the Petersburg academy of science gave China books and astronomical instruments: a refractor telescope from 3 ½ to 5 sazhens; a telescope for studying comets; two English globes: terrestrial and celestial; an English pocket chronometer; 2 or 3 portable barometers and a thermometer; books on astronomy in English and French; the complete collection of the Petersburg almanac for 1809-1819.[3]

In the second half of the 19[th] century, the Imperial Academy of Sciences arranged to build a magneto-meteorological observatory[4] in Beijing, the first director of which was Konstantin Skachkov[5], a disciple of the astronomists A.N. Savich and V.Ya. Struve.

Skachkov spent a total of 25 years in China, and learned Chinese so well that he could study ancient Chinese manuscripts on economics, politics and astronomy[6]. He received a considerable amount of literature as a gift from members of the religious mission, and also from Chinese scholars and officials. He received several rare and valuable works on Chinese astronomy from the uncle of the Chinese emperor, who was interested in Skachkov's activities at the observatory attached to the religious mission. This uncle was the chairman of the astronomical tribunal, but as Skachkov discovered, he did not even know arithmetic, and Skachkov became his teacher. On his return to Russia, Skachkov brought a library of Chinese manuscripts with him, which consisted of 11,000 volumes (they became part of the collection of the Rumyantsev Museum[7]), and wrote several works on the history of Chinese astronomy. In 1874, his study "The fate of astronomy in China" was published [Skachkov].

Skachkov was familiar with ancient Chinese treatises and the history of astronomy, had studied manuscripts left by the Jesuits, and was familiar with the level of Chinese astronomers, and he harshly criticized the picture of ancient Chinese astronomy given by the Jesuits, accusing them of falsifying information about China, from erroneous maps to praising Chinese astronomy.

---

[2] Iakinf (Bichurin) was the head of the 9[th] religious mission from 1808 to 1820. It was one of the most brilliant in the entire history of the mission, and its pupils and Bichurin himself made a significant mark on international Sinology.

[3] St. Petersburg Branch of the RAS Archive. F.2., Op.1. 1819. D. 1. On instructions for the mission travelling to China. L. 70-71.

[4] In 1848, construction of the Magneto-meteorological observatory began at the Russian religious mission in Beijing. The observatory also conducted astronomical studies, and was included in the network of physical observatories of Russia [8, p. 33]

[5]. Konstantin Andrianovich Skachkov (1821-1883) – interpreter for the Russian Ministry of Foreign Affairs, diplomat and sinologist. Disciple of the astronomers A.N. Savich and V.Ya. Struve. He considered I. Bichurin to be his tutor. When he returned home, Skachkov already possessed a large collection of Chinese manuscript books, maps and albums of drawings. In 1866 he taught Chinese at the St. Petersburg University. On Skachkov, see http://drevlit.ru/docs/kitay/XIX/1840-1860/Skackov/biogr.php The Skachkov collection at the manuscript section of the Russian State Library (f. 273) contains two manuscript volumes "Materials for studying Chinese astronomy" (615 leaves). He should not be confused with the Sinologist Pyotr Yemelyanovich Skachkov (1892–1964), who nevertheless wrote this biographical study of the work of K.A. Skachkov.

[6] It was very difficult to translate Chinese books on astronomy. Skachkov's correspondent was a Sinologist, Stanislav Julienne from France. He wrote that he would rather translate dozens of pages from the Chinese classics than a few lines from their astronomy. In the list of his translations, K.A. Skachkov wrote in brackets next to "History of Astronomy of the Han Dynasty": "Unfinished because of incomprehensibility".

[7] It is now held in the archives of the Lenin State Library.

Skachkov writes that the observatory built by the Jesuits stands unused, the Chinese do not know how to use its instruments; mathematical and astronomical books and tables left by the Jesuits lie in storage, and they are sold for a pittance at book stalls, but do not find buyers. "The more I read, the more I immersed myself in the Chinese language, and the study of Chinese astronomy, the more I became convinced that the entire path taken in China by astronomy from ancient times does not at all confirm the reputation of Chinese astronomers, but rather shows their hopelessness; but on the other hand, this survey of Chinese astronomy, from ancient times, is an interesting contribution to the history of science. Throughout the entire 19[th] century, there was no serious voice in literature among Sinologists about Chinese astronomy". [Skachkov, p. 11]. Skachkov notes that ancient manuscripts mention teachers and pupils who came from afar, and who imparted their knowledge. For example, concerning the calendar, Skachkov writes: "experts on celestial science were sent for. These experts were found, and thanks to them the Chinese introduced a calendar based on the system of the music of the spheres. In imitation of sounds heard in nature, bells were introduced for the first time, and later they were replaced by bamboo flutes. According to the length of the pipes, the sounds they made determined the nature of the four seasons; and the remultiplication of mutual relations in the length of these pipes determined the sizes expressed in the units of the day, for the periods of the monthly movement of the moon, the year and so on." [Skachkov, p. 14–15]. Skachkov believes that these teachers were foreigners, firstly Babylonians, then Greeks, Indians and Arabs. Skachkov notes the similarity of the ancient Chinese tales of the sky with myths of different people, and also the utilitarian requirements of astronomy in China throughout its history – fortune-telling practices etc., the ability to predict eclipses, and the need for a calendar. He established that over 1,500 years in China, 220 names of astronomers (astrologers) were recorded, and around 900 works on the science of the heavens were written during this time. Skachkov takes into account the mention of their names, preserved fragments, and citations in other treatises. In the library of the imperial palace, Skachkov read 74 works about the sky, and he contributed 76 Chinese books on astronomy to the collection of the Rumyantsev museum. Analyzing these works, Skachkov concludes that the achievements of Chinese astronomy are expressed in "more or less precise definitions of the annual movement of the sun and moon and its anomalies, classification of the planets Jupiter and Saturn, detecting the angle of obliquity of ecliptic, determining the periods of return of eclipses, observing and determining the prediction of the spring equinox etc. But astrological requirements, which have long had an honorary place in their calendars, led the Chinese to study the starry sky, with determinations of the position of stars; this led to the geographical determination of the latitude and longitude of many points in China". Skachkov says that the reason for the lack of theoretical astronomy is that "theoretical philosophers interfered in Chinese astronomical literature, who lacked the most basic knowledge of the rudiments of the study of the sky, and took obscure lines by ancient sages and various sophisms to ascribe laws to celestial motion and phenomena, leading real researchers astray". At the same time, Skachkov pays his respects to the Chinese astronomers who preserved notes about astronomic phenomena from the second century BC. "We may see that Chinese astronomy from the 2[nd] century BC and later resembles Greek astronomy: and then from the 9[th]-17[th] century, Chinese astronomy is similar, and sometimes even identical to the astronomy of the Indians and the Arabs" [Skachkov, p. 25]. The conclusions that Skachkov draws are the following: "1) The Chinese did not create their astronomy themselves, but borrowed it from foreigners. 2) In the subsequent development of their astronomy, the Chinese used manuals of the Greeks, Indians, Arabs and Europeans. 3) At present, Chinese knowledge of astronomy is at a lower level than it was in Europe before the time of Tycho de

Brage. 4)With the blind, infantile attachment of the Chinese to their classic works, which slow down every step towards progress, with their notable indifference to everything that happens outside their country, and with their inability to study higher mathematics, they will not be able to understand our knowledge of astronomy in the near future. And finally, 5) the study of the numerous monuments of astronomical literature preserved in Chinese writing since times of great antiquity may represent a very respectable contribution to science" [Skachkov, p. 31].

*1920, G.N. Popov*. In 1920, the book by G.N. Popov[8], "History of Mathematics" was published, in which he critically analyzes European historical-mathematical literature, including literature on the history of Chinese astronomy. He singles out the 1852 work by A. Wylie, "Jottings on the Science of the Chinese", as the only work which has full and reliable information about the success of the Chinese in mathematics. It was found that long before the 1st century AD, they knew arithmetic and geometry, could solve equations, and judging by surviving treatises, created their own techniques, and by the 3rd century BC also showed incredible skill in solving complex issues of indeterminate equations. They could inevitably encounter difficulties in the simplest tasks connected with compiling the calendar, for example in determining a number which when divided by certain numbers would have given certain remainders. <…> A number of studies mentioned the methods of Chinese mathematicians which they used in indeterminate analysis. As we know, similar problems were solved by the Indians, but their methods differed from the Chinese. The Chinese in the 8th century AD solved problems which were solved in the 19th century by Gauss and Dirichlet." [Popov, p. 233–234]. Popov shows the external influences of other cultures: from the 3rd century BC to the 1st century AD – Buddhism, from the 7th century – Islam, Nestorian Christianity, Manicheism, and from the 13th century Catholicism. "The consequences of this synthesis, it seems likely to us, had fateful significance for Chinese science. In general, with little inclination towards the methodological and systematic treatment of ideas in the field of precise knowledge, the Chinese were simply overwhelmed by the mass of new ideas. We believe that a mentally stronger nation would also have found itself in a difficult position". [Popov, p. 236]. His full historical survey ends with the words: "We have seen that the latest studies have nullified the importance of Chinese astronomy as a science." [Popov, p. 233].

*1934. A.V. Marakuev. "The sources of ancient astronomy".* The Soviet Oriental scholar Alexander Vladimiriovich Marakuev[9] (1891–1955) was the first to notice the emergence of two opposing tendencies in views on the history of Chinese astronomy. "For an illustration of the diametrically opposed viewpoints on the ancient astronomy of China, I can refer on the one hand to the author of the article Astronomy in the 14th edition of the Encyclopedia Britannica (1925), which supports the hypothesis of the gray antiquity of Chinese astronomy. In his statements he relies, also without reservations, on examples that have been questioned many times and are still suspect: the solar eclipse which Chinese court astronomers were allegedly supposed to predict (but did not predict), calculations in the 22nd century BC and the precise determination of the inclination of the ecliptic, for which the Chinese obtained the value of $23^0$ 54'3" in the 12th century BC. This value was given in Shu-qing, almost exactly coinciding with the calculations by Laplace.

On the other hand, the renowned Japanese Sinologist Iijima Tadao[10] (飯島忠夫) claimed in his works that Chinese astronomy began, at the earliest, in the 4[th] century BC. And it evidently emerged under the influence of Greek culture, which was brought to China in the era that immediately followed Alexander the Great's campaign in India (330 BC)" [A.V. Marakuev]. In his paper, Marakuev notes the important influence of intercultural ties in the development of astronomical knowledge in China, gives a periodization of the history of Chinese astronomy, and provides parallels between Chinese astronomical knowledge and those in other cultures. Marakuev notes that in 104 BC, when the Taichuli calendar was adopted, a period of 135 months was used for the cycle of eclipses, while in the West a cycle of 223 months was known (the Chaldean Saros). This is an argument in favor of the independence of Chinese astronomy from western influence. Marakuev also examines the historians' dispute on the origin of the system 28 lunar houses (star groups) that was widespread in antiquity. There are two opinions on the place of its origin: Babylon and China. Marakuev concludes: "At present I agree with Iijima on the more likely western origin of Far Eastern astronomy; however, before the 2[nd] century BC we do not see direct traces of western influence, but it is possible that they were brought in from the outside, without leaving written traces. But this does not mean that the issue is solved, and that we will not examine it again when we have gathered more material" [Marakuev].

*1980. E.I. Beryozkina "Mathematics of Ancient China" and other works*. Elvira Ivanovna Beryozkina (born 1931) translated into Russian "Mathematics in Nine Books" [Mathematics in Nine Books] – an Ancient Chinese treatise on mathematics. This was the first complete translation into a European language. She made studies on Ancient Chinese mathematics, which were included in her book. As she writes in the introduction, "mathematics in China developed from deep antiquity more or less independently, and achieved its greatest development in the 14[th] century AD. Subsequently, western mathematics entered China, mainly brought by European missionaries, and this already marks a different era in the history of science in China. The ancient tradition in mathematics was thus broken and lost. Many discoveries made earlier than they were in Europe were forgotten, and they were repeated once more by western scientists. Chinese mathematicians, studying ancient and medieval texts, found incredible results obtained by their ancestors. In this book, the main attention is given to mathematics of ancient China in the period from the 2[nd] century BC to the 7[th] century AD. Works by Chinese mathematicians of the 13[th]-14[th] centuries are examined in less detail". [Beryozkina, Mathematics of Ancient China, p. 3].

In 1995-2003 Yao Fang[11], a post-graduate student at Moscow State University, made a partial translation of the "Treatise on the Gnomon" (Zhoubi suanjing[12]) with commentary by Zhao Qunqing (ca. 3[rd] century BC) from Ancient Chinese into Russian [Yao Fang].

*2009 The Religious Culture of China: an encyclopedia in 5 volumes*. Recently a wonderful encyclopedia of the religious culture of China was published. [The religious culture of China: an encyclopedia in 5 volumes. V. 5. Science, technical and military thought, health and education / Edited by M.L. Titarenko, A.I. Kobzev, V.Ye. Yeremeev, A.E. Lukyanov. Moscow: Oriental literature RAS, 2009]. The fifth volume contains articles by V.Ye. Yeremeev[13]. "Mathematics", "Astronomy", "The Calendar" and many others, with a broad survey of the history of achievements of Chinese science, and an analysis of its interaction with other civilizations. In our opinion, the author managed to avoid favoring either polar viewpoint on the origin of Chinese astronomy, and his articles give an objective outline of historical information. His posthumously published article "Science in the Yuan and Ming Dynasties"[14] (2012) is also worthy of attention, which give a full picture of the development of astronomy in this period.

Russian sinology reached its zenith in the late 19[th] – first third of the 20[th] century, when a large number of literary monuments of the history of China were studied, translated and published. But after K.A. Skachkov and the few researchers named above, no one in Russia paid attention to the history of Chinese astronomy.

In 1930, the Stalinist repressions weakened the scientific potential of Russian sinologists, many of whom were executed, or received long prison sentences. For many decades, Sinology was restricted by official ideology. The study of rebellious movements, the history of the Chinese Communist Party and folk culture came to the fore.

In the second half of the 20[th] century, a new generation of Russian researchers revived Sinology in the scientific centers of Moscow, Leningrad and Vladivostok. In 1983-1984, the "Great Chinese-Russian Dictionary:" was published in 4 volumes (1983-1984), and the encyclopedia "The Spiritual Culture of China" in 6 volumes[15] (2006–2010).

Of subsequent major publications, we may name the first complete translation of Sima Qian. Historical Notes. Shiqi. In 9 volumes / Translation from the Chinese, introduction, commentary and appendix by R.V. Vyatkin. Moscow: Nauka, 1972-2010.

Despite the impressive body of Russian sinology studies, in studying the history of Chinese astronomy, only K.A. Skachkov (1874), A.V. Marakuev (1934), Yao Fang (1995–2003) and V.Ye. Yermeev (2009, 2012) went directly to primary sources.

We note the chief difficulties of the historiography of ancient Chinese astronomy: widespread dubious dating of primary sources, frequent interpretation of ancient models in terms of modern science, insufficient study of ancient intercultural ties and influences[16]. Transcription of Chinese names and titles in different European systems also creates difficulties. A further problem is the separation of the history of ancient Chinese astronomy itself from the history of astrology, and its subordinate role in state ideology. We should add that the translation and interpretation of ancient Chinese astronomical texts is not just difficult for philologists, but for astronomers as well. These problems require a balanced and cautious attitude to assessments.